\documentclass{rspublic}
\usepackage{amsmath}
\usepackage{color,graphicx}
\begin{document}

\title[GRBs: Cosmic Rulers?]{Gamma Ray Bursts: Cosmic Rulers for the High-Redshift Universe?}
\author[F.~C.~Speirits, M.~A.~Hendry, A.~Gonzalez]{Fiona C. Speirits$^1$, Martin A.
Hendry$^1$, Alejandro Gonzalez$^{1,2}$} \affiliation{1. Dept.~of
Physics \& Astronomy, University of Glasgow, Glasgow G12 8QQ, UK. 2.
Division Academica de Ciencias Basica, Universidad Juarez Autonoma
de Tabasco, Km. 1 Carr. Cunduacan, C.P. 86690, Tabasco, Mexico}

\label{firstpage} \maketitle

\abstract{The desire to extend the Hubble Diagram to higher
redshifts than the range of current Type Ia Supernovae observations
has prompted investigation into spectral correlations in Gamma Ray
Bursts, in the hope that standard candle-like properties can be
identified. In this paper we discuss the potential of these new
`cosmic rulers' and highlight their limitations by investigating the
constraints that current data can place on an alternative
Cosmological model in the form of Conformal Gravity. By fitting
current Type 1a Supernovae and Gamma Ray Burst (GRB) data to the
predicted luminosity distance redshift relation of both the standard
Concordance Model and Conformal Gravity, we show that currently
\emph{neither} model is strongly favoured at high redshift. The
scatter in the current GRB data testifies to the further work
required if GRBs are to cement their place as effective probes of
the cosmological distance scale.}{cosmology; gamma ray bursts;
supernovae; concordance model; conformal gravity}

\section{The Ghirlanda Relation as a Cosmic Ruler}
Recently a number of authors have highlighted the potential of
long-duration Gamma Ray Bursts (GRBs) as distance indicators. The
most promising indicator appears to be the so-called `Ghirlanda
Relation' (Ghirlanda et al. 2004): the tight correlation between the
isotropic equivalent energy and the peak energy of the GRB
integrated spectrum.

However, several authors (e.g. Friedman \& Bloom 2004) have pointed
out potential sources of systematic error which may undermine the
application of the Ghirlanda Relation as a distance indicator, for
example Dai et al.~(2004) assume a cosmology when calibrating the
relation, which presents a circularity issue when using it to fit
cosmological parameter values.

Friedman \& Bloom conclude that the Ghirlanda Relation provides no
significant improvement in the constraints on $\Omega_M$ and
$\Omega_{\Lambda}$. In their view, this is mainly due to the
currently small number of GRB calibrators, including the lack of
low-redshift GRBs. Contributions to the uncertainty also arise from
the sensitivity to data selection choices and to the values and
ranges assumed for the number density of the surrounding medium and
the efficiency of each event.

Notwithstanding the caveats of Friedman \& Bloom, we have recently
considered their application to test the viability of Conformal
Gravity theories (Mannheim 2003). Mannheim's theory makes a
specific, and very strong, prediction: the expansion of the Universe
has always been accelerating. However, the Hubble Diagram for this
model does not diverge from the corresponding Friedmann model until
$z>1$.

In Mannheim's Conformal Gravity theory the luminosity distance
redshift relation is given by
\begin{equation}
\label{cgdl}
d_L=-\frac{c(1+z^2)}{H_0q_0}\left[1-\left(1+q_0-\frac{q_0}{(1+z)^2}\right)^{1/2}\right]
\end{equation}

\noindent where $z$ is the redshift of the source, $H_0$ is the
Hubble parameter, $c$ is the speed of light and $q_0$ is the model
deceleration parameter, related to the Concordance Model parameters
by $q_0=\frac{\Omega_M}{2}-\Omega_{\Lambda}$

\section{Results and Conclusions}

\begin{figure}[h!]
\centering
\begin{minipage}[c]{.5\linewidth}
\centering
\includegraphics[width=\linewidth]{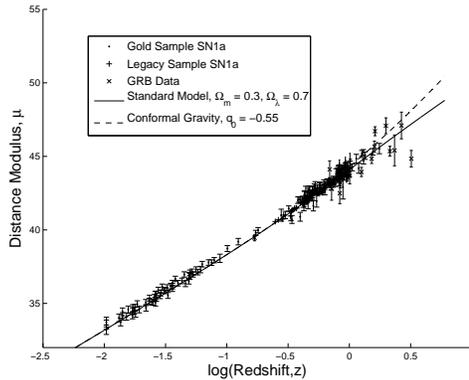}
\end{minipage}%
\begin{minipage}[c]{.4\linewidth}
\centering \renewcommand{\figurename}{Fig.}\caption{The log(z)-$\mu$
Hubble Diagram for Standard Model and Conformal Gravity with SN and
GRB data sets} \label{fig}
\end{minipage}
\end{figure}

We have used data on 150 Gold Sample Type 1a Supernovae from Riess
et al.~(2004), 71 SN from the first year results of the Supernova
Legacy Survey (Astier et al. 2006) and 19 GRBs compiled by Friedman
and Bloom (2004), employing a cut at $cz<5000kms^{-1}$ to remove the
effect of peculiar velocities from the SN data. The Hubble diagram
for these data sets can be seen in Fig.~\ref{fig}. We have compared
these data with distance moduli predicted for the Standard Model
with ($\Omega_M=0.3$, $\Omega_{\Lambda}=0.7$) and the corresponding
Conformal Gravity model with $q_0=-0.55$. Values for
$\sigma_{\mu_{obs}}$ for our SN were taken from the published data
source, while for the GRBs they were calculated following Dai et
al.~(2004). These fits give $\chi^2_{/d.o.f}=4.91$ and 5.65
respectively.

From our results, we see that -- with current SN + GRB data -- the
specific prediction of Mannheim's Conformal Gravity that the
universe did not undergo a deceleration phase remains viable.
However, the large $\chi^2$ per degree of freedom for \emph{both}
the Conformal Gravity and Friedmann models shown in Fig.\ref{fig}
should sound an important note of caution regarding the efficacy of
GRBs as distance indicators. The GRB data do not yet appear good
enough to reliably discriminate between models which accelerate and
decelerate above $z=1$.

\section*{References}

Astier,~P.,~et al. 2006 The Supernova Legacy Survey: Measurement of
$\Omega_M$, $\Omega_{\Lambda}$ and $w$ from the First Year Data Set.
\emph{A\&A} \textbf{447}, 31-48.\\
Dai,~Z.~G., Liang,~E.~W., Xu,~D., 2004 Constraining $\Omega_M$ and
Dark Energy with Gamma-Ray Bursts. \emph{ApJ} \textbf{612},
L101-L104.\\
Friedman,~A.~S. and Bloom,~J.~S., 2005 Toward a More Standardized
Candle Using Gamma-Ray Burst Energetics and Spectra. \emph{ApJ}
\textbf{627} 1-25.\\
Mannheim,~P.~D., 2003 Options for Cosmology at Redshifts Above One.
In \emph{AIP Conf. Proc. 672: Short Distance Behavior of Fundamental
Interactions} (ed. Kursunoglu~B.~N. and Camcigil,~M. and
Mintz,~S.~L. and Perlmutter,~A.) pp.~47-64.\\
Riess,~A.~G.,~et al. 2004 Type Ia Supernova Discoveries at $z>1$
from the Hubble Space Telescope: Evidence for Past Deceleration and
Constraints on Dark Energy Evolution. \emph{ApJ} \textbf{607},
665-687.
\end{document}